# The Impossible Trio in CDO Modeling

Emmanuel Schertzer
+1 212 412 2888
emmanuel.schertzer@barcap.com

Yadong Li
+1 212 412 6869
yadong.li@barcap.com

Umer Khan
+44 20 313 40258
umer.khan@barcap.com

**We show that stochastic recovery always leads to counter-intuitive behaviors in the risk measures of a CDO tranche - namely, continuity on default and positive credit spread risk cannot be ensured simultaneously. We then propose a simple recovery variance regularization method to control the magnitude of negative credit spread risk while preserving the continuity on default.**

## 1. Introduction

The base correlation model with Gaussian copula is the standard model for synthetic CDOs. Earlier generations of the base correlation model assume the recovery rate to be deterministic. Although this assumption is appealing in its simplicity, it is problematic when trying to calibrate market quotes in the context of elevated systemic risk. Indeed, in a deterministic recovery setting, super senior tranches are always risk free. However, since the beginning of the credit crisis in 2007, the market has priced in significant systemic risk, and observed super senior tranche prices are anything but risk free. Therefore, models with deterministic recovery assumptions can no longer calibrate to market tranche prices, and stochastic recovery models have been adopted in order to calibrate the base correlation model to market tranche prices.

(Amraoui and Hitier 2008) and (Krekel 2008) are two of the representative stochastic recovery models in which recovery rates depend on the common market factor, so realized recovery rates are likely to be smaller if there are more defaults. This dependency between the recovery rate and the common market factor has been successful in generating a level of risk for super senior tranches comparable with what is observed in the market. However, the risk measures of models with stochastic recovery often exhibit unintuitive behaviors in practice, including discontinuity on default, negative credit spread risk, etc. We examine the root causes of these unintuitive behaviors and show that they are inevitable in any stochastic recovery model.

Section 2 provides an argument for the previous claim. More precisely, it shows that no model accounts for the price of the risky super-senior tranche while 1) being continuous on default and 2) exhibiting positive credit spread risk. Section 3 illustrates this result with examples. Sections 2 and 3 suggest that any model for pricing CDOs is likely to be imperfect and that a trade-off has to be made. Section 4 proposes a simple method to control the negative spread risk while preserving the continuity on default.

## 2. The Impossible Trio

We first introduce some terminology that will be used in further discussions. Consider a CDO tranche consisting of N names indexed by i = 1…N. We denote by $s_i$ the spread of the i-th name. For the sake of simplicity, we assume that credit curves are flat, but our arguments and conclusions can easily be extended to cover shaped curves. Adopting standard terminology, the market recovery of a name will refer to the (risk-neutral) expected value of the recovery given default. In particular, we note that in a constant and deterministic recovery setting, market recovery and realized recovery always coincide. Next, we will say that a tranche is super senior whenever its attachment point is greater than the total portfolio loss assuming that all the names in the portfolio default at their market recoveries.[1] For example, a 60-100% tranche is super senior if all the names' market recoveries are greater or equal to 40%. As already noted, super senior tranches are risk free under the deterministic recovery assumption by definition, implying that a stochastic recovery assumption must be adopted in order to mirror the level of riskiness observed in today's market. Finally, we define two fundamental risk measures for the i-th name in the portfolio. Denoting by $PV(s_i)$ the current present value (PV) and assuming the spread of name i to be a variable $s_i$ (all other spreads in the portfolio are treated as constant), we define:

- $\text{CreditSpread01}_i = \dfrac{\partial PV(s_i)}{\partial s_i}$,

---

[1] The 15-30% CDX tranche is often referred to as the super senior tranche, but it is not "super senior" in our definition.

  which is the PV change of the tranche due to a small change in the name's credit spread.

- $VOD(s_i) = PV(D_i) - PV(s_i)$

  The value-on-default (VOD) is the PV change of the tranche if the name suddenly jumps to default with the market recovery rate[2]. The symbol $D_i$ is used to represent the default state for name i. Note that in the definition of the VOD, $PV(D_i)$ includes the tranche's protection payment (if any) from the default event and the adjustments to the tranche notional and strikes by removing the defaulted name from the underlying portfolio.

Here, the VOD definition is limited to the case in which the realized recovery is identical to the market recovery rate before the default event. In reality, the realized recovery can be different from the market expected recovery. The PV difference caused by the recovery change is defined as a separate risk measure called Recovery01. The overall default event can then be viewed as two steps: 1) a default at market recovery and 2) a change from market recovery to realized recovery. The VOD risk as defined captures the PV change from the first step, and the Recovery01 risk captures the PV change from the second step.[3]

**Natural properties of the risk measures:** Having introduced the main risks in CDO pricing, we explain the desirable properties of an ideal CDO model. (As discussed later, these natural properties are unfortunately mutually exclusive.) We always consider the risk of a tranche from the protection buyer's point of view.

First, CreditSpread01 is expected to be positive, since the total expected portfolio loss should increase with any individual credit spread. From a hedging point of view, a negative CreditSpread01 is problematic since it requires a long protection tranche to be hedged by buying additional protection on the underlying single names, which is rather counter-intuitive. It is, therefore, highly desirable for a CDO model to ensure the positivity of CreditSpread01s.

Another desirable property is continuity on default. A model is said to be continuous on default if

$$\lim_{s_i \to \infty} VOD(s_i) = PV(D_i) - PV(s_i = \infty) = 0$$

We used $s_i = \infty$ to represent the state in which a name's spread is so high that its default is certain and imminent. Continuity on default implies that the value of a tranche does not change if a certain and imminent default occurs at the market recovery rate.

---

[2] In practice, there are several weeks between the default event (e.g., a company files bankruptcy or misses a debt payment) and the settlement of the defaulted name in derivative contracts. Only on the settlement date is the tranche's protection payment made and the defaulted name removed from the underlying portfolio. In this article, we ignore the difference between the default date and settlement date and assume that default and settlement happen at the same time.
[3] The recovery rate change of a defaulted but unsettled name will change the tranche PV because it affects the tranche's protection payment, strikes, and notional amount after the settlement.

We note that the stochastic recovery assumption often leads to discontinuity on default. Indeed, if an imminent default is fully anticipated by the market, the only new information from the actual default at the market recovery rate is the removal of recovery uncertainties (or recovery variance). In other words, the recovery changes from a stochastic state to a purely deterministic state. Since CDO tranches are sensitive to the variance of the underlying portfolio loss, this collapse of recovery variance explains the discontinuity on default in many stochastic recovery models (for instance, the model described in [Amraoui & Hitier 2008]).

Discontinuous-on-default models create problems in managing default event risk because of the difficulty in hedging the PV jump from the collapse of the recovery variance. Typical hedging instruments, such as single-name CDSs or corporate bonds, are continuous on default, so they cannot be used to hedge such valuation jump. As discussed earlier, the overall PV change from a default event can also include the Recovery01 contribution from the difference between realized recovery and market recovery; in the definition of continuity in default, we are not concerned about the Recovery01 portion of the PV change, since it can be effectively hedged by CDS or corporate bonds.

Continuity on default is another highly desirable model property since it removes the un-hedgable valuation jumps due to recovery variance changes. As discussed in Sections 3 and 4, the key to building a continuous-on-default model is forcing the recovery variance to zero when a name approaches default, thus removing the variance changes from the default event. (Bennani & Maetz 2009) is an example of such a continuous-on-default stochastic recovery model.

Finally, we note that the reduction in recovery variance for very risky names actually agrees with market reality, since more information about an issuer's business and balance sheet becomes public as it approaches default. Therefore, market participants generally have a much better reading of the recovery rate for distressed issuers than for highly rated issuers.

**The impossible trio:** So far, we have described three highly desirable properties of a CDO model:

1. Super senior tranche can be risky (i.e., with positive protection value)
2. Always positive CreditSpread01
3. Always continuous on default

The question is whether any model can satisfy all three properties. Unfortunately, the properties are not compatible with each other and, thus, cannot co-exist in *any* CDO model. This argument is based on the idea that statements 1, 2, and 3 imply that there exists a zero coupon tranche containing an underlying name whose VOD has the following three properties simultaneously:

i. $VOD(s_i) < 0$ for some values of $s_i$

ii. $VOD(s_i)$ monotonically decreases with $s_i$

iii. $\lim_{s_i \to \infty} VOD(s_i) = 0$

Obviously, i, ii, and iii cannot be true simultaneously, since a negative, decreasing, and continuous function can never go back to zero as $s_i \to \infty$. This implies that the three desired model properties (1, 2, and 3) cannot co-exist either.

We now turn to the justification of i, ii, and iii from properties 1, 2, and 3:

**3➔iii**: This is the definition of continuity on default.

**2➔ ii**: This is due to the following relationship between CreditSpread01 and VOD:

$$\text{CreditSpread01}_i = -\frac{\partial VOD(s_i)}{\partial s_i} \tag{1}$$

which directly follows the definition of CreditSpread01 and VOD by noticing that the $PV(D_i)$ (the PV of a tranche after defaulting the name i) does not depend on the defaulted name's spread $s_i$ prior to the default event. Indeed,

$$\begin{aligned}\frac{\partial VOD(s_i)}{\partial s_i} &= \frac{\partial}{\partial s_i}(PV(D_i) - PV(s_i)) \\ &= -\frac{\partial PV(s_i)}{\partial s_i} \\ &= -\text{CreditSpread01}_i\end{aligned}$$

where the second equality follows from the independence of the defaulted PV with respect to $s_i$.

**1➔i**: This is intuitive if we consider an investor buying protection on a zero-coupon risky super senior tranche whose PV is the same as its protection value. Consider a scenario in which all the names in the tranche portfolio default sequentially at the market expected recovery. Under this scenario, each default event will change the tranche PV by the defaulting name's VOD.[4] The total PV change from defaults in all the names is, therefore, the sum of all the intermediate VODs. On the other hand, the investor started with some positive protection value in the super senior tranche; but the tranche protection becomes worthless after all the names default at their market-expected recoveries. Therefore, all the intermediate VODs must sum up to the net negative PV change to the investor, so at least one of the intermediate VODs is negative[5].

We stress that no specific model assumptions have been made and that our results are generic and apply to all CDO models. We refer to the three desirable model properties as the "impossible trio." The best a CDO model can achieve is two out of three. Among the impossible trio, today's market conditions demand the risky super senior tranche; therefore, practitioners have to choose between positive CreditSpread01 and continuity on default when building CDO models.

## 3. Recovery Variance Regularization

We use the stochastic recovery model suggested in (Andersen and Sidenius 2004) to illustrate the trade-off between positive CreditSpread01 and continuity on default. (Andersen and Sidenius 2004) models the expected recovery rate conditioned on the common factor z:

$$R_i(t, z) = \mathbb{E}[r_i | \tau_i < t, z] = R_m \Phi(\alpha z + \beta_i(t)) \tag{2}$$

---

[4] The intermediate VODs are with respect to the tranche that has experienced all the previous default events instead of the original tranche.
[5] Here, the existence of negative VOD is shown only for a zero-coupon super senior tranche, but in practice the negative VODs often occur in non-super senior and couponed tranches as well.

where $\Phi()$ is the cumulative normal distribution function, $\tau_i$ is the default time of the name, $r_i$ is the recovery rate given default, $\alpha$ and $R_m$ are constants, and $\beta_i(t)$ is determined by matching the expected recovery of the name's CDS:

$$\mathbb{E}[\mathbf{1}_{\tau_i<t}r_i] = \mathbb{E}[\mathbb{E}[\mathbf{1}_{\tau_i<t}|z]R_i(t,z)] = \mathbb{E}[p_i(t,z)R_i(t,z)] = p_i(t)R_i \qquad (3)$$

where $R_i$ is the market recovery rate for name i, $p_i(t)$ is the name's default probability at time t, and $p_i(t,z) = \mathbb{E}[\mathbf{1}_{\tau<t}|z]$ is the default probability conditioned on market factor z. The constant parameter $R_m \in [R_i, 1]$ is the maximum possible value for $R_i(z,t)$ and also controls the recovery rate variance: the greater $R_m$, the greater the recovery variance. We note that when $R_m = R_i$, the model reverts to deterministic recovery. Indeed,

$$\mathbb{E}[p_i(t,z)R_i(t,z)] = \mathbb{E}[p_i(t,z)R_i\Phi(\alpha z + \beta_i(t))] \leq \mathbb{E}[p_i(t,z)R_i] = p_i(t)R_i,$$

implying that the only way to satisfy (3) is to choose $\beta_i(t) = \infty$, thus having $R_i(z,t) = R_i$. We use this extensively in the remainder of this section.

**Figure 1: Recovery Variance**

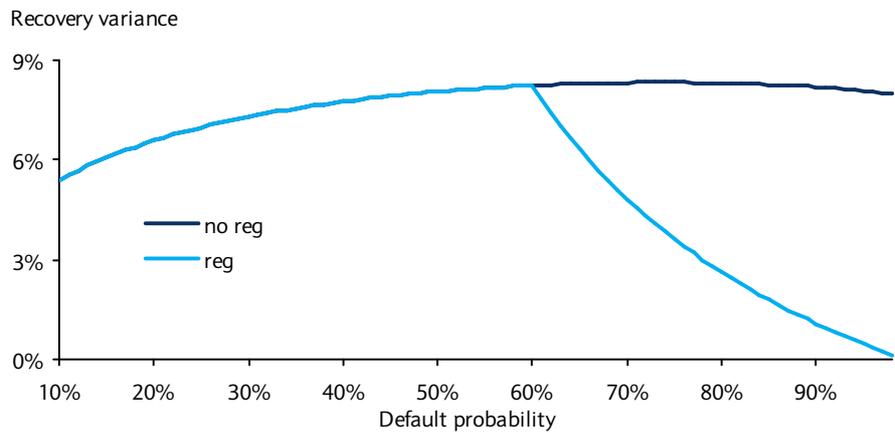

Source: Barclays Capital

In light of Section 2, it is natural to ask whether the recovery model in (2) is continuous on default or produces positive credit spread risk. The answer lies in the choice of parameter $R_m$. We examine three cases.

$R_m = 1$: In this case, the stochastic recovery model is not continuous on default because the variance of the recovery does not go to zero even when a name is close to default, as shown by the top line in Figure 1. The bottom line in Figure 2 is a representative $VOD(s_i)$ function obtained using the stochastic recovery model in (2) for a long protection 15-30% CDX-IG9 tranche as we increase a name's credit spread. We can see that it is a monotonically decreasing function of spread; therefore, CreditSpread01 remains positive according to (1).

$\lim_{p_i \to 1} R_m(p_i) = R_i$: It is easy to modify the previous recovery model to make it continuous on default. We make $R_m$ in (2) a continuous function of the default probability $p_i(t)$ and force it to approach $R_i$ when $p_i(t) \to 1$. This effectively forces the model to revert to deterministic recovery for very risky names. We use the term "recovery variance regularization" to refer to this technique of forcing an existing stochastic recovery model's recovery variance to zero as $p_i(t) \to 1$ in order to ensure continuity on default. The bottom line in Figure 1 shows an example of recovery variance regularization in which the recovery variance is forced to decrease as the name's default probability goes over 60%. The center line in Figure 2 is the corresponding $VOD(s_i)$. As can be seen from the chart, the recovery variance regularization is effective in making the model continuous on default. Also, notice that the center line starts to increase when spread is at roughly 2000bp, implying that the recovery variance regularization starts to produce negative CreditSpread01 for names with spreads higher than 2000bp.

$R_m = R_i$: For comparison, Figure 2 also shows the $VOD(s_i)$ for the same tranche using a deterministic recovery model (top line). It shows that the deterministic recovery model is continuous on default and has positive CreditSpread01s for all spread levels. However, as mentioned earlier, this model fails to assess the riskiness of the super-senior tranche.

**Figure 2: VOD from Different Stochastic Recovery Models**

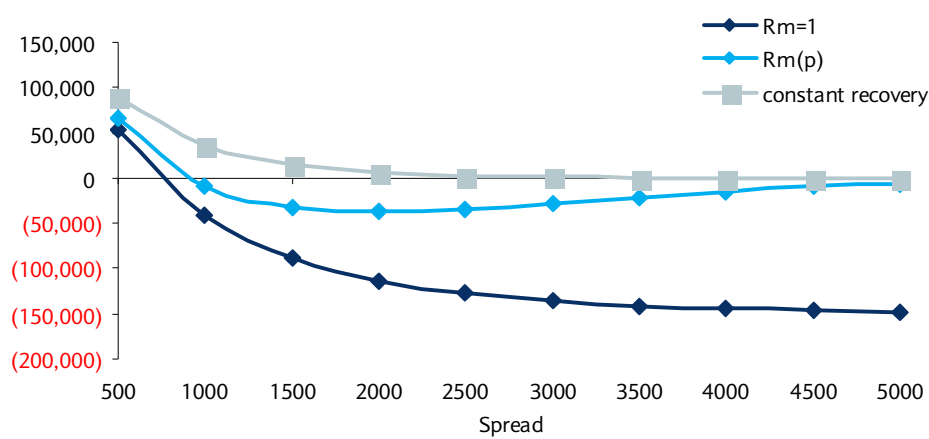

Source: Barclays Capital

**Figure 3: Summary of Recovery Model Properties**

| Recovery Models | Allow Risky Super Senior Tranche | Always Positive CreditSpread01 | Always Continuous on Default |
|---|---|---|---|
| Deterministic Recovery | No | Yes | Yes |
| Stochastic Recovery w/o Variance Regularization | Yes | Yes | No |
| Stochastic Recovery with Variance Regularization | Yes | No | Yes |

Source: Barclays Capital

Figure 3 summarizes the three different recovery models we have considered, each of which satisfies two properties of the impossible trio. So far, we have considered only the recovery models; other modeling assumptions could also preclude certain desired properties. Using the standard base correlation model as an example, negative CreditSpread01 can arise from the base correlation mapping or interpolation, as discussed in (Morgan and Mortensen 2007). Certain base correlation mapping methods (such as the tranche loss proportion mapping) can also cause discontinuity on default. Please refer to (Baheti & Morgan 2007) for a more detailed discussion of mapping methods.

Among the impossible trio, we argue that it is better to choose the continuity on default than the always positive CreditSpread01s for the following reasons:

- Recovery variance regularization can be chosen so that it creates additional negative CreditSpread01s only for very high spread names. In practice, when a name becomes very risky, it is more important to manage its default and recovery risk than spread risk. Therefore, the benefits from being continuous on default outweigh the problems from negative CreditSpread01.

- The reduction in recovery variance when a name approaches default agrees with market observations.

- Positive CreditSpread01 is not always preserved under base correlation interpolation or mapping; hence, it is difficult to remove negative CreditSpread01s completely even when giving up the continuity on default. In contrast, the continuity-on-default property is more robust and can be ensured by using recovery variance regularization and a mapping method that is known to be continuous on default (for example, probability matching mapping) under the base correlation model.

Recovery variance regularization is an easy and effective way to retrofit an existing stochastic recovery model with the continuity on default property; it is also efficient numerically, since there is no additional computational cost from the original stochastic recovery model described earlier with $R_m = 1$.

Because of the generality of the impossible trio, more sophisticated recovery models, such as the spot recovery model considered in (Bennani & Maetz 2009), face exactly the same trade-off between continuity on default and positive CreditSpread01. We think the benefit of a more sophisticated recovery model relative to a simple model such as (2) with recovery variance regularization is questionable for pricing CDOs, since the added numerical complexity does not translate into significantly better model properties.

## 4. Choosing $R_m(p)$

As discussed earlier, we want to calibrate the function $R_m(p)$ so that the recovery variance starts to collapse only when the default probability reaches a certain threshold. When default probability is low, we want to keep $R_m(p) = 1$ so that it preserves the behavior of the original stochastic recovery model in (2). The key question is how to choose the threshold: if it is too low, safer names will be assigned negative CreditSpread01 and the uncertainty on recovery will be insufficient to calibrate super-senior tranches; if it is too high, the VOD will revert to zero abruptly, producing strong gamma and very negative CreditSpread01 for risky names.

We provide some heuristic arguments on how to choose the best functional form of $R_m(p)$. The key observation is that in a continuous-on-default model, negative CreditSpread01s are direct results of the negative VODs. Once the VOD becomes negative, it has to increase in order to go back to zero when $s_i \to \infty$, resulting in negative CreditSpread01s according to (1). Therefore, if one can reduce the magnitude of negative VODs, it would also confine the magnitude of negative CreditSpread01s.

As proved in Section (2), negative VOD is a consequence of the riskiness of super senior tranches. The payoff function of a super senior tranche with strike K is given as:

$$U(L) = (L - K)^+ = (\sum_i l_i - K)^+$$

where $L = \sum_i l_i$ is the total portfolio loss and $l_i$ is the loss of individual names. If we only consider name i's loss $l_i$ and hold all other names' losses constant, the tranche payoff function $U(L)$ is a positive, increasing, and convex function of $l_i$. The $U(L)$ function is difficult to work with analytically. Instead, we consider the simplest positive, increasing, and convex payoff function in $l_i$: $W(l_i) = l_i^2$. To ensure that the VOD will always be positive for the payoff function $W(l_i)$, we must have:

$$\bar{V}OD = \mathbb{E}[W(1 - R_i)] - \mathbb{E}[W(\mathbf{1}_{\tau_i \leq t}(1 - r_i))] \geq 0 \qquad (4)$$

The payoff function $W(l_i)$ is very tractable analytically, and we can derive a function form of $R_m(p)$ so that (4) is always true. In the appendix, we prove the following result:

**Proposition 1.** *Under the model described in (2), choosing*

$$R_m(p_i) = 1 - (1 - R_i)\max(0, \frac{R_i - (1 - p_i)}{R_i p}) \qquad (5)$$

*ensures the positivity of the VOD function under the payoff function $W(l_i) = l_i^2$. Remarkably, the result is independent of the parameter $\alpha$ in (2).*

We then argue that the $R_m(p)$ in (5) would also limit the magnitude of the negative VOD for the actual tranche payoff function $U(L)$. Figure 4 shows the $R_m(p_i)$ function in (5) for the 40% market recovery rate. Note that the recovery variance starts to decrease when the default probability reaches the threshold of $1 - R_i$. Even though the argument leading to (5) is mostly heuristic, it works well in practice, as it strikes the right balance between the needs to keep safer names' CreditSpread01s positive and to control the negative CerditSpread01 and gamma effect for risky names.

## 5. Conclusion

The three desirable model properties we call the impossible trio cannot co-exist in any CDO model. Given that, we argue that it is better to choose continuity on default than always positive CreditSpread01. A simple recovery variance regularization technique is an effective and numerically efficient way to retrofit any stochastic recovery model with the continuity-on-default property. We use an heuristic argument to determine the simple $R_m(p)$ function in (5), which we think is a good compromise.

**Figure 4: The $R_m(p_i)$ Function**

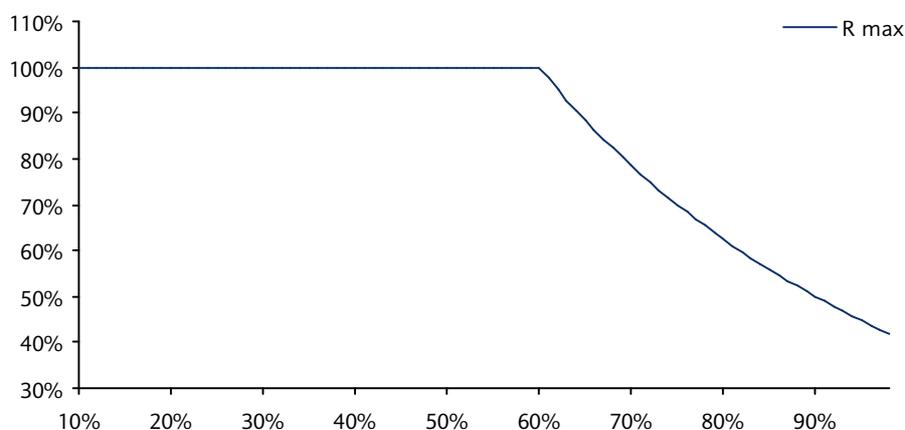

Source: Barclays Capital

## References

Amraoui and Hitier (2008), "Optimal Stochastic Recovery for Base Correlation," *defaultrisk.com*.

Andersen and Sidenius (2004), "Extensions to the Gaussian Copula: Random Recovery and Random Factor Loadings," Jun, Journal of Credit Risk

Baheti and Morgan (2007), "Base Correlation Mapping," Lehman Brothers: Quantitative Credit Research Quarterly

Bennani and Maetz (2009), "A Spot Stochastic Recovery Extension of the Gaussian Copula," defaultrisk.com

Durrett (2004), "Probability: Theory and Examples" Cambridge Press, 2004.

Krekel (2008), "Pricing Distressed CDOs with Base Correlation and Stochastic Recovery," defaultrisk.com

Morgan and Mortensen (2007), "CDO Hedging Anomalies in the Base Correlation Approach," Lehman Brothers: Quantitative Credit Research Quarterly

## Appendix A. Proof of Proposition 1

Since the simplified VOD function defined in (4) depends on a single name, we drop the i index in the formulas to ease the notation. Before giving a proof of our proposition, we return to the (Andersen and Sidenius 2004) construction discussed in Section 3. Recall that the stochastic recovery model $r$ is a random variable satisfying the following conditional expected value condition:

$$\mathbb{E}[r|\tau < t, z] = R(t,z) = R_m \Phi(\alpha z + \beta(t))$$

If one is interested only in a single maturity t (with no time consistency requirement), such a model can be constructed by assuming that

$$r = R_m \Phi(\alpha z + \beta) \qquad (6)$$

Since our proposition deals only with a fixed maturity, we assume without loss of generality that $r$ is defined according to the previous equation. Finally, before the actual proof, we note that $\beta$ is defined as an implicit function of $R_m, \alpha$ through the CDS consistency condition

$$\mathbb{E}(\mathbf{1}_{\tau(\omega,z)<t} r(z)) = pR \qquad (7)$$

where p is the probability of default before time t and R the market recovery of the name under consideration. As a consequence, if one defines the simplified version of the VOD for the underlying name as in (4), the VOD can be seen as a function of our initial choice for $R_m$ and $\alpha$, i.e.

$$VOD \equiv VOD(R_m, \alpha)$$

We will show that the choice of $R_m^*$ in (5) ensures that

$$\forall \alpha, \quad VOD(R_m^*, \alpha) \geq 0$$

in two steps:

Claim 1: We will prove that for any choice of $R_m \in [R_i, 1]$ we have:

$$VOD(R_m, \alpha) \geq \lim_{\alpha \uparrow +\infty} VOD(R_m, \alpha)$$

Claim 2: We will show that

$$lim_{\alpha \uparrow \infty} VOD(R_m^*, \alpha) \geq 0$$

**Proof of Claim 1**

We assume that $R_m$ is fixed and that the definition of our recovery model is only $\alpha$ dependent. To ease the notation, we write

$$X(\alpha) = \mathbf{1}_{\tau_i(\omega, z) < t}\, r(z) \quad \text{and} \quad VOD(R_m, \alpha) \equiv VOD(\alpha) \tag{8}$$

From (6) and (7), $\{X(\alpha)\}_\alpha$ defines a family of random variables satisfying two properties:

$$\begin{aligned} E(X(\alpha)) &= pR \\ 0 \leq X(\alpha) &\leq R_m \end{aligned} \tag{9}$$

To prove our result, we rely on the following lemmas.

**Lemma 1**

$$\forall \alpha_1, \alpha_2, \quad VOD(\alpha_1) - VOD(\alpha_2) = Var(X(\alpha_2)) - Var(X(\alpha_1))$$

Proof. In order to make the dependence of the recovery in $\alpha$ explicit, we denote by $r_\alpha$ the stochastic recovery with underlying parameter $\alpha$.

Using $\Delta VOD$ for the LHS of the identity in our lemma, we have

$$\begin{aligned} \Delta V &= [(1-R)^2 - \mathbb{E}(1_{\tau(\omega,z)<t}(1-r_{\alpha_1}(z))^2)] - [(1-R)^2 - \mathbb{E}(1_{\tau(\omega,z)<t}(1-r_{\alpha_2}(z))^2)] \\ &= [\mathbb{E}(X(\alpha_2)^2) - \mathbb{E}(X(\alpha_1)^2)] - 2[\mathbb{E}(X(\alpha_2)) - \mathbb{E}(X(\alpha_1))] \\ &= Var(X(\alpha_2)) - Var(X(\alpha_1)) \end{aligned}$$

where the first equality follows from the definition of the simplified VOD (4) and the last equality follows from the fact that the two random variables have the same mean.

Q.E.D.

**Lemma 2.** *There exists a real number $K_r$ such that*

$$\begin{aligned} X(\infty) &= 0 \quad \forall z < K_r \\ &= R_m \quad \forall z > K_r \end{aligned}$$

Proof. Recall that in the Gaussian copula framework,

$$\mathbb{P}(1_{\tau<t}|z) = \phi(\sqrt{\tfrac{\rho}{1-\rho}}(K_p - z)) \quad \text{with} \quad K_p = \tfrac{c}{\sqrt{\rho}} \tag{10}$$

where $\rho$ is the correlation between the names in the portfolio and $c = \phi^{-1}(p)$. Then define the number $K_r \in [-\infty, K_p]$ implicitly through the following formula

$$\mathbb{P}(z \in [K_r, K_p]) = \tfrac{R}{R_m} p.$$

(It is easy to show that such a $K_r$ always exists.)

Next, let $\beta = -K_r \alpha$ and define the stochastic recovery using the resulting $\alpha, \beta, R_m$ parameters, i.e.,

$$r(z) = R_m \phi(\alpha(z - K_r)) \tag{11}$$

As $\alpha$ goes to infinity, (10) becomes

$$\begin{aligned} \mathbb{P}(1_{\tau < t} | z) &\approx 1 \quad \text{when} \quad z < K_p \\ &\approx 0 \quad \text{otherwise} \end{aligned} \tag{12}$$

and (11) becomes

$$\begin{aligned} R &\approx R_m \quad \text{when} \quad z > K_r \\ &\approx 0 \quad \text{otherwise} \end{aligned} \tag{13}$$

From there, one can also easily check that the consistency condition (7) is satisfied at the limit.

$$\mathbb{E}(1_{\tau < t} \, r) = \mathbb{E}(1_{z \in [K_r, K_p]}) R_m = R_m \cdot p \frac{R}{R_m} = pR$$

**Q.E.D**

**Lemma 3.** *Let a,b,m be three real numbers and let P be the space of all random variables taking value in the interval [a,b] such that*

$$\forall X \in P, \quad E(X) = m.$$

*Then*

$$\max_{X \in P} Var(X) = Var(\bar{X})$$

*where $\bar{X}$ is the element of P whose mass is concentrated on the extremities of the interval [a, b].*

**Proof.** If one considers only the subset of two point masses in the family P (i.e., a probability distribution concentrated on two points), then the maximum variance is obviously attained for $\bar{X}$. The result is easily extended to any type of r.v. by using the fact that any random variable with mean m can be decomposed into a convex combination of two independent point masses probability distributions with mean m (see Durett [2004] for a proof of this result).

**Q.E.D**

We are now ready to prove claim 1. As mentioned previously, $\{X(\alpha, z)\}_\alpha$ defines a family of random variables with equal mean that are always bounded by 0 and $R_m$. According to Lemma 2, the element of this family with the highest variance is the one with probability mass concentrated on the extreme values zero and $R_m$. By Lemma 2, this is achieved by the element $X(\infty)$. Lemma 1 yields the result.

**Proof of Claim 2**

We use the notation introduced in the proof of Lemma 2 and assume that $\alpha = +\infty$. In particular, the conditional probability (resp., the recovery) is a step function equal to 1 (resp. $R_m$) on $[-\infty, K_p]$ (resp., $[K_r, \infty]$) and zero otherwise (see Equations (12) and (13) in Lemma 2). This implies that the simplified VOD in (4) writes

$$
\begin{aligned}
VOD &= (1-R)^2 - \mathbb{E}(1_{\tau<t}(1-r)^2) \\
&= (1-R)^2 - \mathbb{E}(\mathbb{E}(1_{\tau<t}|z)(1-r(z))^2) \\
&= (1-R)^2 - \mathbb{E}(1_{z<K_p}(1_{z>K_r}(1-R_m)^2 + 1_{z<K_r})) \\
&= (1-R)^2 - (1-R_m)^2 \mathbb{P}(z \in [K_r, K_p]) - P(z < K_r) \\
&= (1-R)^2 - (1-R_m)^2 \mathbb{P}(z \in [K_r, K_p]) - (\mathbb{P}(z<K_p) - \mathbb{P}([z \in [K_r, K_p]))
\end{aligned}
$$

In Lemma 2 (see (11) and (12)), recall that $K_p, K_r$ are defined in such way that

$$\mathbb{P}(z \in [K_r, K_p]) = p\frac{R}{R_m} \quad \text{and} \quad \mathbb{P}(z < K_p) = p$$

Implying that

$$VOD = (1-R)^2 - (1-R_m)^2 p\frac{R}{R_m} - p(1 - \frac{R}{R_m})$$

Finally, one can directly check that taking $R_m$ as in (5) yields a positive VOD by plugging (5) into the latter formula.

**Q.E.D.**